\documentclass[%
 reprint,
 amsmath,amssymb,
 aps,
]{revtex4-1}

\usepackage{graphicx}
\usepackage{dcolumn}
\usepackage{bm}

\begin{document}

\preprint{APS/123-QED}

\title{Perturbation, Non-Guassianity and Reheating in a GB-$\alpha$-Attractor Model }

\author{Kourosh Nozari}
 \homepage{knozari@umz.ac.ir}
\author{Narges Rashidi}
\homepage{n.rashidi@umz.ac.ir}%
\affiliation{Department of Physics, Faculty of Basic Sciences,\\
University of Mazandaran,\\
P. O. Box 47416-95447, Babolsar, IRAN}

\date{\today}

\begin{abstract}
Motivated by $\alpha$-attractor models, in this paper we consider a
Gauss-Bonnet inflation with an E-model-type of potential. We consider
the Gauss-Bonnet coupling function to be the same as the E-model
potential. In the small $\alpha$ limit we obtain an attractor at
$r=0$ as expected, and in the large $\alpha$ limit we recover the
Gauss-Bonnet model with potential and coupling function of the form
$\phi^{2n}$. We study perturbations and non-Gaussianity in this
setup and we find some constraints on the model's parameters in
comparison with PLANCK data sets. We study also the reheating epoch
after inflation in this setup. For this purpose, we seek the number
of e-folds and temperature during reheating epoch. These quantities
depend on the model's parameter and the effective equation of state
of the dominating energy density in the reheating era. We find some
observational constraints on these parameters.
\begin{description}
\item[PACS numbers]
04.50.Kd , 98.80.Cq , 98.80.Es
\item[Key Words]
Inflation, Cosmological Perturbations, Non-Gaussianity, Reheating,
$\alpha$-Attractor, Observational Constraints
\end{description}
\end{abstract}

\maketitle


\section{Introduction}

Inflation is accepted as a testable paradigm for the physics of the
early Universe. This scenario addresses some important
problems of the standard big bang cosmology satisfactorily. The most important
problem in standard cosmology is the origin of the cosmological
structures. In this regard, the most favorite inflationary models
are those models in which the needed primordial density perturbations,
seeding the structure formation in the Universe, can be generated naturally
~\cite{Gut81,Lin82,Alb82,Lin90,Lid00a,Lid97,Rio02,Lyt09}. In the
simplest realization of the inflation, the Universe is dominated by
a nearly flat potential energy of a single scalar field (the
inflaton field). Quantum fluctuations of the scalar field during the
inflationary epoch are considered as the primordial source of the
cosmological perturbations. The dominant mode of the primordial
density perturbations in the simple single field model (with
canonical kinetic term) is almost adiabatic and scale invariant.
Also, the distribution of the perturbations for this simple case is predicted to be
Gaussian~\cite{Mal03}. Nevertheless, with the advancement of
technology, the observational cosmologists have shown that the
primordial density perturbations are somehow scale
dependent~\cite{Ade15a,Ade15b}. Despite this fact that currently
there is no \emph{direct} signal for non-Gaussian feature of the
perturbations, Planck collaboration has obtained some constraints on
the primordial non-Gaussianity~\cite{Ade15c}. In this regard, by
proposing extended inflationary models, some authors have tried to
realize theoretically a level of non-Gaussianity in the dominant
mode of the primordial density perturbation~\cite{Mal03,Bar04,Che10,Fel11a,Fel11b,Noz12,Noz13a,Noz13b,Noz13c,Noz16a,Bab04a,Che08,Noz15,Bart10,Fasi14}.
Carrying a large amount of information about the cosmological
dynamics by the non-Gaussian perturbations, makes the models predicting non-Gaussianity more interesting.

Incorporating some quantum corrections into Einstein gravity seems
to be necessary when we study the very early Universe (towards the Planck scale).
In fact, a low-energy limit of a quantum theory
of gravity is the Einstein gravity. It is believed that a
promising candidate for quantum gravity is the string theory. The Gauss-Bonnet (GB) term is the simplest
higher-order curvature correction to the gravitational action that is motivated by string models.
Including the GB term as $R_{abcd}R^{abcd}-4R_{ab}R^{ab}+R^{2}$ in the theory, which has an
important role in the dynamics of the early
Universe~\cite{Zwi85,Bou85}, leads to a ghost-free action and also
there is no problem with the unitarity. However, it turns out that
in dimensions less than five, the Gauss-Bonnet (GB) term has no
influence on the background dynamics and is just a topological term.
In this regard, one has to consider this term in higher
dimensions~\cite{Bro07,Bam07,And07,Noz08,Noz09a,Noz09b,Noz09c}. On
the other hand, to see the effects of the Gauss-Bonnet term in the
background dynamics in 4-dimension, we should consider a function of
GB term or couple this term nonminimally to a scalar filed~\cite{Noj05,Noj07,Guo09,Guo10,Satoh08a,Satoh08b,Hikm16}. We note that the GB term appears to be coupled to the dilaton field in string theory in a particular form. For instance in Refs.~\cite{Bam07,Guo07,Iih11}, the authors have chosen an exponential coupling as $e^{-\lambda\phi}$ where $\lambda$ is a constant. Here we consider a coupling function that has the potential to recover the well-known coupling of the dilaton to the GB term in string theory. On the other hand, scalar fields and also field-dependent couplings are present usually in the string effective action. These scalar fields are moduli associated to geometrical properties of compactified extra dimensions. Our motivation here is to extend the inflaton field to be one of these moduli fields. We have made the assumption that the inflaton field, as a scalar field responsible for cosmological inflation, belongs to the dilaton fields family.

Recently, the idea of ``cosmological attractors'' in the inflation models
has attracted much attention. Among the models incorporating the idea of
cosmological attractors, we can refer to conformal
attractors~\cite{Kal13a,Kal13b} and $\alpha$- attractors
models~\cite{Kai14,Fer13,Kal13c,Kal14}. For more details on the issue of $\alpha$- attractor
see also Refs.~\cite{Cec14,Kal14b,Lin15,Jos15a,Jos15b,Kal16,Sha16,Odi16}. The
conformal attractor models have the characteristic property that in
large $N$ limit, the spectral index of the primordial
curvature perturbations and the tensor-to-scalar ratio always are given as
$n_{s}=1-\frac{2}{N}$ and $r=\frac{12}{N^{2}}$ respectively. For a single field
$\alpha$-attractor models, in the small $\alpha$ limit we have
$n_{s}=1-\frac{2}{N}$ and $r=\frac{12\alpha}{N^{2}}$. In this paper,
we are interested in $\alpha$-attractor models. There are two types
of this models called E-model and T-model. The E-model is specified
by the potential as
\begin{equation}
\label{eq1}
V=V_{0}\Big[1-\exp\big(-\sqrt{\frac{2\kappa^{2}}{3\alpha}}\phi\big)\Big]^{2n}\,,
\end{equation}
and the T-model is specified by
\begin{equation}
\label{eq2}V=V_{0}\tanh^{2n}\Big(\frac{\kappa\phi}{\sqrt{6\alpha}}\Big)\,,
\end{equation}
where $V_{0}$, $n$ and $\alpha$ are some free parameters. In the large $\alpha$ limit, the predictions of these
models converge towards the predictions of the simple single field
inflation model with the potential as $V\sim\phi^{2n}$. In this
paper we are interested in E-models of $\alpha$-attractors and study
primordial perturbations and non-Gaussianinty in this class of
models.

An important issue in cosmological inflation is the
reheating process after the end of inflation. As long as the
slow-roll conditions ($\eta,\epsilon\ll 1$) (requiring the potential
to be sufficiently flat) are satisfied, the Universe inflates. As
soon as the slow-roll conditions break down, the inflaton rolls to
the minimum of its potential and starts to oscillate. In a simple
canonic reheating scenario, this oscillation of the inflaton, by
processes including the physics of particle creation and
non-equilibrium phenomena, causes the inflaton decay to the plasma
of the relativistic particles leading to the radiation-dominated
Universe~\cite{Ab82,Do82,Al82}. Nevertheless, some other complicated
scenarios of reheating have been proposed which include also the
non-perturbative processes. Among them we can refer to instant
preheating~\cite{Fel99}, the parametric resonance
decay~\cite{Ko94,Tr90,Ko97} and tachyonic
instability~\cite{Gr97,Sh06,Du06,Ab10,Fel01a,Fel01b}. The reheating
era, is characterized by parameters $N_{rh}$ (number of e-folds
during reheating) and $\omega_{eff}$ (the effective equation of
state parameter during reheating). In this regard, studying
$N_{rh}$, $\omega_{eff}$ and also $T_{rh}$ (the reheating
temperature) in the inflation models seems to be useful to constrain
these models~\cite{Dai14,Un15,Co15,Cai15,Ue16}. 
See also Ref.\cite{Amin15} for a recent and elegant review on reheating.

In this paper, we consider an inflation model in which the scalar
field is nonminimally coupled to the Gauss-Bonnet (GB) term in
4-dimension. The potential of the scalar field is considered to be E-model
$\alpha$-attractor. We also assume that the coupling function
between the scalar field and the GB term is the same as the
potential, that is,
\begin{equation}
\label{eq3}{\cal{G}}(\phi)={\cal{G}}_{0}\Big[1-\exp\big(-\sqrt{\frac{2\kappa^{2}}{3\alpha}}\phi\big)\Big]^{2n}\,.
\end{equation}
In this regard, when $\alpha\rightarrow \infty$, the GB coupling
function approaches to ${\cal{G}}(\phi)\sim{\cal{G}}_{0}\phi^{2n}$
and when $\alpha\rightarrow 0$, the GB term has no effect on the
dynamics. In section II, by obtaining the main equations, we study
the background dynamics of this GB-$\alpha$-attractor model. In
section III we study the linear perturbations (including the
scalar and tensor perturbations) in our setup and we obtain
expressions for the scalar spectral index and tensor-to-scalar ratio
in terms of $\alpha$ and ${\cal{G}}_{0}$. Then we perform the
numerical analysis on the model's parameter space. In section VI,
the non-linear perturbations and non-gaussian feature of perturbations distribution are
considered in details. We obtain the amplitude of the non-Gaussianity in terms
of the sound speed (which is a function of the model's parameters) and in the equilateral limit of the momenta. Then we study the
evolution of $f_{NL}^{equil}$ versus $c_{s}^{2}$ numerically. After
that, in section V, we study the reheating epoch in this
GB-$\alpha$-attractor model. In this section we calculate
expressions for the number of e-folds and temperature during the
reheating phase after inflation. In this regard we obtain more
constraints on the model's parameters. Finally, in section VI, the
summary of our analysis is presented.

\section{The Setup}
The action for an inflation model including a nonminimally coupled Gauss-bonnet term in
4-dimensions is written as
\begin{eqnarray}
\label{eq4} S=\int
d^{4}x\sqrt{-g}\Bigg[\frac{1}{2\kappa^{2}}R-\frac{1}{2}\partial_{\mu}\phi
\partial^{\mu}\phi-V(\phi)-{\cal{G}}(\phi){\cal{L}_{GB}}
\Bigg],\nonumber\\\hspace{0.6cm}
\end{eqnarray}
where, $R$ is the Ricci scalar, $\phi$ is an inflaton filed and
$V(\phi)$ is the E-model potential defined in equation \eqref{eq1}.
Also, ${\cal{G}}(\phi)$ is the Gauss-Bonnet coupling term given by
equation \eqref{eq3}. ${\cal{L}_{GB}}$ is the lagrangian of the
Gauss-Bonnet term. For a spatially flat FRW geometry, Friedmann equation of this model is given as follows
\begin{equation}
\label{eq5}
H^{2}=\frac{\kappa^{2}}{3}\left(\frac{1}{2}\dot{\phi}^{2}+V(\phi)+24H^{3}\dot{{\cal{G}}}\right),
\end{equation}
where a dot marks a time derivative of the corresponding
parameter. The following equation of motion
\begin{eqnarray}
\label{eq6} -\ddot{\phi}-3H\dot{\phi}
-24H^{4}{\cal{G}}'-24H^{2}\dot{H}{\cal{G}}'-V'=0\,\hspace{0.6cm}
\end{eqnarray}
is obtained by varying the action \eqref{eq4} with respect to the
inflaton field and a prime marks differentiation with respect to $\phi$.

The definition of the slow-roll parameters as
$\epsilon\equiv-\frac{\dot{H}}{H^{2}}$ and
$\eta=-\frac{1}{H}\frac{\ddot{H}}{\dot{H}}$, gives the following
expressions for $\epsilon$ and $\eta$ in terms of the model's
parameters
\begin{eqnarray}
\label{eq7} \epsilon=-{\frac {{\cal{F}}}{ \left( 2\,H -24\,
H^{2}{\kappa}^{2}\dot{{\cal{G}}} \right) H^{2}}}\,,
\end{eqnarray}
\begin{eqnarray}
\label{eq8} \eta=-{\frac {\dot{{\cal{F}}}}{H \,{\cal{F}} }}+{\frac {2\,\dot{H} -48
\,H {\kappa}^{2} \dot{{\cal{G}}}\, \dot{H} -24\,
H^{2}{\kappa}^{2}\ddot{{\cal{G}}} }{H\left( 2\,H -24\,
H^{2}{\kappa}^{2}\dot{{\cal{G}}} \right) }}\,,
\end{eqnarray}
where the parameter ${\cal{F}}$ is defined as follows
\begin{eqnarray}
\label{eq9}{\cal{F}}\equiv 8H^{3}{\kappa}^{2}\ddot{{\cal{G}}}+\frac{\kappa^{2}}{3}\,
V' \dot{\phi} +\frac{\kappa^{2}}{3}\,\dot{\phi}\, \ddot{\phi}\,.
\end{eqnarray}

Since the Hubble parameter evolves during the inflationary era
so slowly, the conditions $\epsilon \ll 1$ and
$\eta \ll 1$ are satisfied in this period. When one of these parameters reaches 
unity, the inflation phase terminates automatically.

Although in a simple single field model the slow-roll limits are
defined as $\ddot{\phi}\ll |3H\dot{\phi}|$ and $\dot{\phi}^{2}\ll
V(\phi)$, in the presence of GB term there are more conditions. To
have the slow-roll regime in this model, the conditions
$8\kappa^{2}H|\dot{{\cal{G}}}|\ll 1$ and $|\ddot{{\cal{G}}}|\ll
|\dot{{\cal{G}}}H|$ should also be satisfied
(see~\cite{Noz16a,Guo10}). By regarding these conditions and
considering the slow evolution of the Hubble parameter ($\dot{H}\ll
H^{2}$), we get
\begin{equation}
\label{eq10} H^{2}\simeq \frac{\kappa^{2}}{3}V\,,
\end{equation}
and
\begin{equation}
\label{eq11} -3H\dot{\phi}
-24H^{4}{\cal{G}}'-V'\simeq0\,.\hspace{0.6cm}\,.
\end{equation}
In this model, from the definition of the e-folds number as
\begin{equation}
\label{eq12} N=\int_{t_{hc}}^{t_{f}} H dt\,,
\end{equation}
where $t_{hc}$ and $t_{f}$ are the time of horizon crossing and end of inflation respectively,
and equation \eqref{eq9}, we get
\begin{equation}
\label{eq13} N\simeq \int_{\phi_{hc}}^{\phi_{f}} \frac{-3H^{2}
}{V'+24H^{4}{\cal{G}}'} d\phi\,.
\end{equation}
In this paper, we are going to investigate the cosmological viability of
the GB model with E-model type potential. In this regard, one way is
to study the perturbations and non-Gaussian features of this model.
In the following section, we focus on the linear
perturbations to obtain the tensor and scalar spectral indices and
ratio of their amplitudes.

\section{Linear Perturbations}

By using the Arnowitt-Deser-Misner (ADM) formalism
which is given by the following metric~\cite{Arn60}
\begin{equation}
\label{eq14}
ds^{2}=-N^{2}dt^{2}+h_{ij}\big(dx^{i}+N^{i}dt\big)\big(dx^{j}+N^{j}dt\big),
\end{equation}
where $N^{i}$ is the shift vector and $N$ is the lapse function, one can
expand the action \eqref{eq4} up to the second and third orders
in the small fluctuations of the space-time background metric (which
contribute in the perturbations). By expanding the shift and laps
functions in terms of the 3-scalars ${\cal{R}}$ and ${\cal{Y}}$ and
a vector $v^{i}$ (with condition $v^{i}_{,i}=0$) as $N^{i}\equiv
{\cal{Y}}^{i}=\delta^{ij}\partial_{j}{\cal{Y}}+v^{i}$ and
$N=1+{\cal{R}}$, the general perturbed form of the ADM metric
\eqref{eq14} can be obtained~\cite{Muk92,Bau09}. By defining the
parameters ${\Theta}_{ij}$ as a spatial symmetric and traceless
shear 3-tensor and ${{\cal{D}}}$ as the spatial curvature
perturbation, we can write the coefficient $h_{ij}$ in the second
term of equation \eqref{eq14} as
$h_{ij}=a^{2}\left[(1-2{\cal{D}})\delta_{ij}+2{\Theta}_{ij}\right]$.
Therefore, we can write the perturbed form of the metric
\eqref{eq14} as
\begin{eqnarray}
\label{eq15} ds^{2}=
-(1+2{\cal{R}})dt^{2}+2a(t){\cal{Y}}_{i}\,dt\,dx^{i}\hspace{1.5cm}\nonumber\\
+a^{2}(t)\left[(1-2{{\cal{D}}})\delta_{ij}+2{\Theta}_{ij}\right]dx^{i}dx^{j}.
\end{eqnarray}
Working within the uniform-field gauge (specified by $\delta\phi=0$)
is convenient to study the scalar perturbation of the theory. In
this regard, by adopting ${\Theta}_{ij}=0$ and considering the
scalar part of the metric components, we get the following
expression for the perturbed metric ~\cite{Muk92,Bau09,Bar80}
\begin{eqnarray}
\label{eq16}
ds^{2}=-(1+2{\cal{R}})dt^{2}+2a(t){\cal{Y}}_{,i}\,dt\,dx^{i}\hspace{1.5cm}\nonumber\\
+a^{2}(t)(1-2{{\cal{D}}})\delta_{ij}dx^{i}dx^{j}.
\end{eqnarray}

The second order (quadratic) action is obtained by replacing the
perturbed metric \eqref{eq16} in action \eqref{eq4} and expanding it
up to the second order in small perturbations. The result is as
follows
\begin{eqnarray}
\label{eq17} S_{2}=\hspace{7.5cm}\nonumber\\ \int dt\,d^{3}x\,
a^{3}\bigg\{-3(\kappa^{-2}-8H\dot{{\cal{G}}})\dot{{\cal{D}}}^{2}-\frac{2(\kappa^{-2}-8H\dot{{\cal{G}}})}{a^{2}}{\cal{R}}
\partial^{2}{{\cal{D}}}\nonumber\\
+\frac{1}{a^{2}}\Big[2(\kappa^{-2}
-8H\dot{{\cal{G}}})\dot{{\cal{D}}}-\big(2\kappa^{-2}H-24H^{2}\dot{{\cal{G}}}\big){\cal{R}}\Big]\partial^{2}{\cal{Y}}
+\nonumber\\3\Big(2\kappa^{-2}H-24H^{2}\dot{{\cal{G}}}\Big){\cal{R}}\dot{{\cal{D}}}
-\Big(3\kappa^{-2}H^{2}-\frac{\dot{\phi}^{2}}{2}-48H^{3}\dot{{\cal{G}}}\Big){\cal{R}}^{2}\nonumber\\
+\frac{\kappa^{-2}-8H\dot{{\cal{G}}}}{a^{2}}(\partial{{\cal{D}}})^{2}\bigg\}\,.\hspace{1cm}
\end{eqnarray}
The following expressions are the equations of motion of ${\cal{R}}$
and ${\cal{Y}}$, which are obtained by variation of the quadratic
action \eqref{eq17} with respect to the corresponding parameters
\begin{equation}
\label{eq18}
{\cal{R}}=2\frac{\kappa^{-2}-8H\dot{{\cal{G}}}}{2\kappa^{-2}H-24H^{2}\dot{{\cal{G}}}}\dot{{\cal{D}}},
\end{equation}
\begin{eqnarray}
\label{eq19}
\frac{\partial^{2}{\cal{Y}}}{a^{2}}=3\dot{{\cal{D}}}-\frac{\kappa^{-2}-8H\dot{{\cal{G}}}}{a^{2}(2\kappa^{-2}H-24H^{2}\dot{{\cal{G}}})}
\partial^{2}{{\cal{D}}}\nonumber\\-\frac{2(3\kappa^{-2}H^{2}-\frac{1}{2}\dot{\phi}^{2}
-48H^{3}\dot{{\cal{G}}})}{(2\kappa^{-2}H-24H^{2}\dot{{\cal{G}}})}{\cal{R}}. \nonumber\\
\end{eqnarray}
If we use the equation \eqref{eq18} to replace ${\cal{R}}$ in action
\eqref{eq17} and taking some integration by part, we get (to see more
details on calculation of the second and third order actions, see
Refs.~\cite{Fel11a,Fel11b,Che08,See05,Noz16a})
\begin{equation}
\label{eq20} S_{2}=\int
dt\,d^{3}x\,a^{3}{\cal{W}}_{s}\left[\dot{{\cal{D}}}^{2}-\frac{c_{s}^{2}}{a^{2}}(\partial
{{\cal{D}}})^{2}\right],
\end{equation}
where the parameters ${\cal{W}}_{s}$ and $c_{s}^{2}$(dubbed sound velocity), are functions of the model's parameters as
follows
\begin{eqnarray}
\label{eq21}
{\cal{W}}_{s}\equiv-4\frac{\left(\kappa^{-2}-8H\dot{{\cal{G}}}\right)^{2}\left(9\kappa^{-2}H^{2}-\frac{3}{2}\dot{\phi}^{2}
-144H^{3}\dot{{\cal{G}}}\right)}{3\left(
2\kappa^{-2}H-24H^{2}\dot{{\cal{G}}}\right)^{2}}\nonumber\\
+3\left(\kappa^{-2}-8H\dot{{\cal{G}}}\right),\hspace{1cm}
\end{eqnarray}
and
\begin{widetext}
\begin{eqnarray}
\label{eq22}
c_{s}^{2}\equiv3\Bigg[2\Big(2\kappa^{-2}H-24H^{2}\dot{{\cal{G}}}\Big)\Big
(\kappa^{-2}-8H\dot{{\cal{G}}}\Big)H
-\Big(2\kappa^{-2}H-24H^{2}\dot{{\cal{G}}}\Big)^{2}
\Big(\kappa^{-2}-8H\dot{{\cal{G}}}\Big)^{-1}\Big(\kappa^{-2}-8\ddot{{\cal{G}}}\Big)\nonumber\\
+4\Big(2\kappa^{-2}H-24H^{2}\dot{{\cal{G}}}\Big)
\frac{d\Big(\kappa^{-2}-8H\dot{{\cal{G}}}\Big)}{dt}
-2\Big(\kappa^{-2}-8H\dot{{\cal{G}}}\Big)\,
\frac{d(2\kappa^{-2}H-24H^{2}\dot{{\cal{G}}})}{dt}\Bigg]
\Bigg[ \Bigg(9\Big(2\kappa^{-2}H-24H^{2}\dot{{\cal{G}}}\Big)^{2}\nonumber\\
-4\Big(\kappa^{-2}-8H\dot{{\cal{G}}}\Big)\Big(9\kappa^{-2}H^{2}-\frac{3}{2}\dot{\phi}^{2}
-144H^{3}\dot{{\cal{G}}}\Big) \Bigg)\Bigg]^{-1}\,.
\end{eqnarray}
\end{widetext}

When we survey the perturbations of a model, studying the scalar
spectral index gives some useful information about the primordial
perturbations. To obtain this parameter, we should find the power
spectrum in the model, by calculating the vacuum expectation value
of ${{\cal{D}}}$ as
\begin{equation}
\label{eq23} \langle
0|{{\cal{D}}}(0,\textbf{k}_{1}){{\cal{D}}}(0,\textbf{k}_{2})|0\rangle
=(2\pi)^{3}\delta^{3}(\textbf{k}_{1}+\textbf{k}_{2})\frac{2\pi^{2}}{k^{3}}{\cal{A}}_{s}\,.
\end{equation}
By $\tau=0$, we mean the time of the end of inflationary era. In
equation \eqref{eq23}, the power spectrum ${\cal{A}}_{s}$ is defined
as
\begin{equation}
\label{eq24}
{\cal{A}}_{s}=\frac{H^{2}}{8\pi^{2}{\cal{W}}_{s}c_{s}^{3}}\,.
\end{equation}
Now, we are in the position to obtain the scalar spectral index from
the power spectrum as follows
\begin{eqnarray}
\label{eq25} n_{s}-1=\frac{d \ln {\cal{A}}_{s}}{d \ln
k}\Bigg|_{c_{s}k=aH}\hspace{3.7cm}\nonumber\\=-2\epsilon-\frac{1}{H}\frac{d
\ln (\epsilon-4\kappa^{2}H{\cal{G}}'\dot{\phi})}{dt}
-\frac{1}{H}\frac{d \ln c_{s}}{dt}\,.
\end{eqnarray}
By calculating the value of the scalar spectral index, we can seek
the scale dependence of the primordial perturbations.

As the scalar part of the perturbations, studying the tensor part
also gives some other information about the initial perturbations. In
this regard, we consider the tensor part of the perturbed metric
\eqref{eq15}. We set the shear 3-tensor ${\Theta}_{ij}$ to be as
\begin{equation}
\label{eq26}
{\Theta}_{ij}={\Theta}_{+}\vartheta_{ij}^{+}+{\Theta}_{\times}\vartheta_{ij}^{\times}\,,
\end{equation}
which is written in terms of the two polarization tensors. These polarization tensors
satisfy the reality and normalization
conditions~\cite{Fel11a,Fel11b}. The following expression is the
quadratic action for the tensor mode of the perturbations (namely,
gravitational waves)
\begin{eqnarray}
\label{eq27} S_{T}=\hspace{7.5cm}\nonumber\\ \int dt\, d^{3}x\,
a^{3}
{\cal{W}}_{T}\left[\dot{\Theta}_{+}^{2}-\frac{c_{T}^{2}}{a^{2}}(\partial
{\Theta}_{+})^{2}+\dot{\Theta}_{\times}^{2}-\frac{c_{T}^{2}}{a^{2}}(\partial
{\Theta}_{\times})^{2}\right].\nonumber\\
\end{eqnarray}
The parameters ${\cal{W}}_{T}$ and $c_{T}^{2}$ in the second order
action \eqref{eq27} are expressed as follows
\begin{equation}
\label{eq28}
{\cal{W}}_{T}\equiv\frac{1}{4\kappa^{2}}\left(1-8\kappa^{2}H{\cal{G}}'\dot{\phi}\right)\,,
\end{equation}
\begin{equation}
\label{eq29} c_{T}^{2}\equiv1+8\kappa^{2}H{\cal{G}}'\dot{\phi}\,.
\end{equation}
To obtain the tensor spectral index ($n_{T}$), we
should firstly obtain the amplitude of the tensor perturbations which is given in our
setup as
\begin{equation}
\label{eq30} {\cal{A}}_{T}=\frac{H^{2}}{2\pi^{2}{\cal{W}}_{T}c_{T}^{3}}\,,
\end{equation}
which gives
\begin{equation}
\label{eq31} n_{T}=\frac{d \ln {\cal{A}}_{T}}{d \ln k}=-2\epsilon\,.
\end{equation}
Finally, tensor-to-scalar ratio is obtained as
\begin{equation}
\label{eq32}
r=\frac{{\cal{A}}_{T}}{{\cal{A}}_{s}}\simeq16c_{s}\left(\epsilon-4\kappa^{2}H{\cal{G}}'\dot{\phi}\right)\,.
\end{equation}
This equation shows that the presence of the Gauss-Bonnet coupling modifies the standard consistency
relation which in a canonical single field model is given by
$r=16\epsilon$ (or $r=-8 n_{T}$). We note that in some other inflation models
such as the non-minimal gravitational coupling models and multiple
fields inflation the consistency relation is modified too~\cite{Fel11a,Lan08,Lan09}.

\begin{figure*}
\includegraphics[width=.38\textwidth,origin=c,angle=0]{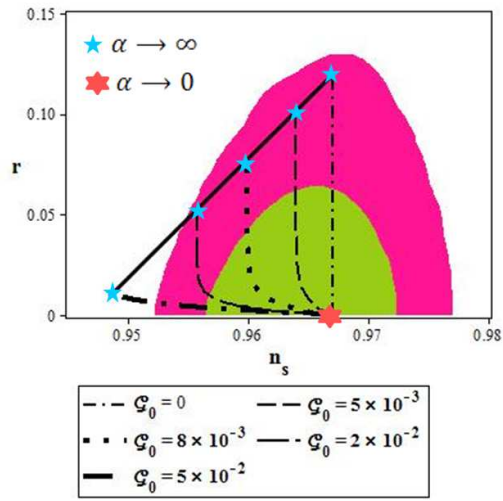}
 \caption{\label{fig1} Tensor-to-scalar ratio versus the scalar
spectral index for a GB-$\alpha$-attractor model in the background
of Planck2015 TT, TE, EE+lowP data. As figure shows, when
$\alpha\rightarrow 0$ all trajectories converge to the point shown
by the red star, corresponding to $r\rightarrow0$ and
$n_{s}\rightarrow 0.967$. Also, when $\alpha\rightarrow\infty$, the
trajectories tend to the points corresponding to the ones with
$V\sim\frac{\phi^{2}}{2}$ and ${\cal{G}}\sim
{\cal{G}}_{0}\frac{\phi^{2}}{2}$. We have chosen $N = 60$ for the number of e-folds of inflation after the relevant CMB scales cross the horizon.}
\end{figure*}

\begin{table*}
\caption{\label{tab:1} The ranges of ${\cal{G}}_{0}$ for which the
values of the scalar spectral index and the tensor-to-scalar ratio
are compatible with Planck2015 observational data. }
\begin{tabular}{cccccccc}
\\ \hline \hline&$\alpha=1$&&$\alpha=10^{2}$&&
$\alpha=10^{3}$&&$\alpha\rightarrow\infty$\\ \hline\\
Planck2015&  ${\cal{G}}_{0}<4.1\times 10^{-2}$
&&${\cal{G}}_{0}<3.91\times 10^{-2}$&& ${\cal{G}}_{0}<2.83\times
10^{-2}$
&&$0.217<{\cal{G}}_{0}<0.41$\\\\
\hline\\\\
\end{tabular}
\end{table*}

For a numerical analysis on the perturbations parameters $n_{s}$ and
$r$, we adopt the potential and the GB coupling function as defined
in equations \eqref{eq1} and \eqref{eq3} (the E-model). In this
regard, we obtain the following expressions
\begin{eqnarray}
\label{eq33}n_{s}=1-\frac{3\,\alpha\, {\cal{V}} ^{2}}{8{n}^{2}{
\kappa}^{4} {\cal{U}} ^{2}}- \Bigg( \frac{\alpha\,\dot{\phi}
{\cal{V}} ^{2}}{8 {n} ^{2}{\kappa}^{4} {\cal{U}} ^{2}}\, \sqrt
{6}\sqrt {{ \frac {{\kappa}^{2}}{\alpha}}}+\frac{\sqrt {6}\alpha\,
{\cal{V}}\dot{\phi}}{8{n}^{2}{ \kappa}^{4}{\cal{U}}} \sqrt {{\frac
{{ \kappa}^{2}}{\alpha}}}\nonumber\\ -{\frac {64}{3\alpha}}\,{
\kappa}^{4}V_{0}{\cal{V}} ^{4n-2} {\cal{G}}_{0}{n}^{2}{\cal{U}}
^{2}\dot{\phi}^{2}+\frac{16}{3\alpha}{\kappa}^{4}V_{0}{\cal{V}}^{4n-1}
{\cal{G}}_{0}n
\dot{\phi}^{2} {\cal{U}}\nonumber\\
+\frac{16}{3\alpha}{\kappa}^{4}V_{0} {\cal{V}} ^{4n-2} {\cal{G}}_{0}
n {\cal{U}} ^{2}\dot{\phi}^{2}-\frac{8}{3}{\kappa}^{2}V_{0}{\cal{V}}
^{4n-1}{\cal{G}}_{0}n\sqrt {6}\sqrt {{\frac
{{\kappa}^{2}}{\alpha}}}{\cal{U}}\ddot{\phi}\Bigg)\times  \nonumber\\
\Bigg(\frac{3\alpha\,{\cal{V}} ^{2}}{16{n}^{2}{\kappa}^{4} {\cal{U}}
^{2}}-\frac{8}{3}{\kappa}^{2}V_{0} {\cal{V}} ^{4n-1}
{\cal{G}}_{0}n\sqrt {6}\sqrt {{\frac {{
\kappa}^{2}}{\alpha}}}{\cal{U}}\dot{\phi} \Bigg) ^{-1}{V_{0}}^{-1}
{\cal{V}} ^{-2n}\nonumber\\-\frac{d \ln c_{s}^{2}}{dt}{V_{0}} ^{-1}
{\cal{V}} ^{-2n}\,,\hspace{1cm}
\end{eqnarray}
\begin{eqnarray}
\label{eq34}r=16c_{s}\bigg[\frac{3\alpha}{16}{\cal{V}}^{2}{n}^{-2}{
\kappa}^{-4} {\cal{U}} ^{-2}-\frac{8}{3}{\kappa}^{2}V_{0} {\cal{V}}
^{4n-1} {\cal{G}}_{0}n \sqrt {{\frac
{6{\kappa}^{2}}{\alpha}}}{\cal{U}}\dot{\phi} \Bigg]\,,\nonumber\\
\end{eqnarray}
where we have defined the parameters ${\cal{V}}$ and ${\cal{U}}$ as
follows
\begin{equation}
\label{eq35}{\cal{V}}=1-{\cal{U}}\,,
\end{equation}
\begin{equation}
\label{eq36}{\cal{U}}={{\rm e}^{-\frac{\sqrt {6}}{3}\sqrt {{\frac {{
\kappa}^{2}}{\alpha}}}\phi }}\,.
\end{equation}

We take $n=1$ and study the evolution of the tensor-to-scalar ratio
versus the scalar spectral index numerically. The results are shown
in figure 1. We have plotted $r$ versus $n_{s}$ for five values of
the nonminimal coupling parameter ${\cal{G}}_{0}$. When
$\alpha\rightarrow 0$ all trajectories converge to the point shown
by the red star, corresponding to $r\rightarrow0$ and
$n_{s}\rightarrow 0.967$. Also, when $\alpha\rightarrow\infty$, the
trajectories tend to the point corresponding to the ones with
$V\sim\frac{\phi^{2}}{2}$ and ${\cal{G}}\sim
{\cal{G}}_{0}\frac{\phi^{2}}{2}$. This GB-$\alpha$-attractor model
is consistent with observational data in some ranges of
${\cal{G}}_{0}$ and ${\alpha}$. In this manner we constrain
${\cal{G}}_{0}$ for some values of ${\alpha}$ in comparison with
Planck2015~\cite{Ade15a} observational data. The results are
summarized in table I.

\section{Nonlinear Perturbations and Non-Gaussianity}
Given that the two-point correlation function of the scalar
perturbations carries no information about the non-Gaussian feature
of the primordial perturbation, to study this feature it is
necessary to calculate the three-point correlation function in our
setup. By expanding the action \eqref{eq4} up to the third order in the small
perturbations, eliminating ${\cal{R}}$ by using the equation \eqref{eq18} and
introducing the auxiliary parameter ${\cal{X}}$ with
\begin{eqnarray}
\label{eq37}
{\cal{Y}}=\frac{2(\kappa^{-2}-8H\dot{{\cal{G}}}){{\cal{D}}}}{2\kappa^{-2}H
-24H^{2}\dot{{\cal{G}}}}
+\frac{a^{2}{\cal{X}}}{\kappa^{-2}-8H\dot{{\cal{G}}}}\,,\hspace{1cm}
\end{eqnarray}
and
\begin{equation}
\label{eq38}
\partial^{2}{\cal{X}}={\cal{W}}_{s}\dot{{\cal{D}}}\,,
\end{equation}
we get
\begin{eqnarray}
\label{eq39} S_{3}=\int dt\, d^{3}x\,\Bigg\{
\Bigg[\frac{3a^{3}}{\kappa^{2}c_{s}^{2}}\,
\Bigg(1-\frac{1}{c_{s}^{2}}\Bigg)
\Bigg(\epsilon-4\kappa^{2}H\dot{{\cal{G}}}\Bigg) \Bigg]{{\cal{D}}}\dot{{\cal{D}}}^{2}\nonumber\\
+\Bigg[\frac{a}{\kappa^{2}}\,\Bigg(\frac{1}{c_{s}^{2}}-1\Bigg)
\Bigg(\epsilon-4\kappa^{2}H\dot{{\cal{G}}}\Bigg)
\Bigg]{{\cal{D}}}\,(\partial{{\cal{D}}})^{2}\nonumber\\+\Bigg[\frac{a^{3}}{\kappa^{2}}\,\Bigg(\frac{1}{c_{s}^{2}\,H}\Bigg)
\Bigg(\frac{1}{c_{s}^{2}}-1\Bigg)\Bigg(\epsilon-4\kappa^{2}H\dot{{\cal{G}}}\Bigg)\Bigg]
\dot{{\cal{D}}}^{3}\nonumber\\-\Bigg[a^{3}\,\frac{2}{c_{s}^{2}}\Bigg(\epsilon-4\kappa^{2}H\dot{{\cal{G}}}\Bigg)\dot{{\cal{D}}}
(\partial_{i}{{\cal{D}}})(\partial_{i}{\cal{X}})\Bigg]\Bigg\}\,.\hspace{0.8cm}
\end{eqnarray}
The 3-point correlators in the interaction picture are obtained as
\begin{eqnarray}
\label{eq40} \langle
{{\cal{D}}}(\textbf{k}_{1})\,{{\cal{D}}}(\textbf{k}_{2})\,{{\cal{D}}}(\textbf{k}_{3})\rangle
=\hspace{3.5cm}\nonumber\\
(2\pi)^{3}\delta^{3}(\textbf{k}_{1}+\textbf{k}_{2}+\textbf{k}_{3}){\cal{B}}_{{\cal{D}}}(\textbf{k}_{1},\textbf{k}_{2},\textbf{k}_{3})\,,
\end{eqnarray}
where
\begin{equation}
\label{eq41}
{\cal{B}}_{{\cal{D}}}(\textbf{k}_{1},\textbf{k}_{2},\textbf{k}_{3})=\frac{(2\pi)^{4}{\cal{A}}_{s}^{2}}{\prod_{i=1}^{3}
k_{i}^{3}}\,
{\cal{N}}_{{\cal{D}}}(\textbf{k}_{1},\textbf{k}_{2},\textbf{k}_{3}).
\end{equation}
The parameter ${\cal{N}}_{{\cal{D}}}$ is defined by the following
expression
\begin{eqnarray}
\label{eq42}
{\cal{N}}_{{\cal{D}}}=\frac{3}{4}\Bigg(1-\frac{1}{c_{s}^{2}}\Bigg)
\Bigg(\frac{2\sum_{i>j}k_{i}^{2}\,k_{j}^{2}}{k_{1}+k_{2}+k_{3}}-\frac{\sum_{i\neq
j}k_{i}^{2}\,k_{j}^{3}}{(k_{1}+k_{2}+k_{3})^{2}}\Bigg)
\nonumber\\-\frac{1}{4}\Bigg(1-\frac{1}{c_{s}^{2}}\Bigg)\Bigg(
\frac{2\sum_{i>j}k_{i}^{2}\,k_{j}^{2}}{k_{1}+k_{2}+k_{3}}-\frac{\sum_{i\neq
j}k_{i}^{2}\,k_{j}^{3}}{(k_{1}+k_{2}+k_{3})^{2}}\hspace{0.5cm}\nonumber\\+\frac{1}{2}\sum_{i}k_{i}^{3}\Bigg)
+\frac{3}{2}\Bigg(\frac{1}{c_{s}^{2}}-1\Bigg)\Bigg(\frac{\left(k_{1}\,k_{2}\,k_{3}\right)^{2}}{(k_{1}+k_{2}+k_{3})^{3}}\Bigg),\hspace{0.5cm}
\end{eqnarray}
by which we can define the so-called ``nonlinearity parameter'',
measuring the amplitude of the non-Gaussianity of the primordial
perturbations, as follows
\begin{equation}
\label{eq43}
f_{NL}=\frac{10}{3}\frac{{\cal{N}}_{{\cal{D}}}}{\sum_{i=1}^{3}k_{i}^{3}}\,.
\end{equation}

The different shapes of the non-Gaussianity are obtained depending
on the different values of the three momenta $k_{1}$ , $k_{2}$ and
$k_{3}$ (for details in this subject
see~\cite{Wan00,Kom01,Bab04,Sen10}). In a simple single field inflation model, where the non-Gaussianity is produced
outside the horizon, the non-Gaussian feature is described with
``local" type (where $k_{3}\ll k_{1}\simeq
k_{2}$)~\cite{Bar04,Kom01,Gan94}. However, there are some models
(such as the DBI, k-inflation and also higher derivative models) where the
non-Gaussianity is created at horizon crossing during inflation. In
these models the bispectrum has a maximal signal when all three
wavelengths are equal to the horizon size ($k_{1}=k_{2}=k_{3}$
)~\cite{Bab04a,Cre06}. In these models when any individual mode is
far outside the horizon, the non-Gaussian feature is suppressed. In
this regard, it is useful to study the non-Gaussianity in the
``equilateral" configuration in these models. So, in our
GB-$\alpha$-attractor model, we study the non-Gaussian feature in
the equilateral configuration. In
this limit we have
\begin{equation}
\label{eq44}
{\cal{N}}_{{\cal{D}}}^{equil}=\frac{17}{72}k^3\left(1-\frac{1}{c_{s}^{2}}\right)\,,
\end{equation}
leading to
\begin{equation}
\label{eq45}
f_{NL}^{equil}=\frac{85}{324}\left(1-\frac{1}{c_{s}^{2}}\right)\,.
\end{equation}

By using this equation and equations \eqref{eq1}, \eqref{eq3} and
\eqref{eq45}, we can perform some numerical analysis on the model's
parameters. The results are shown in figure 2. As this figure shows,
in this GB-$\alpha$-attractor model there is a set of model
parameters which makes $f_{NL}^{equil}$ large enough to be detected
in future surveys.

\begin{figure*}
\includegraphics[width=.38\textwidth,origin=c,angle=0]{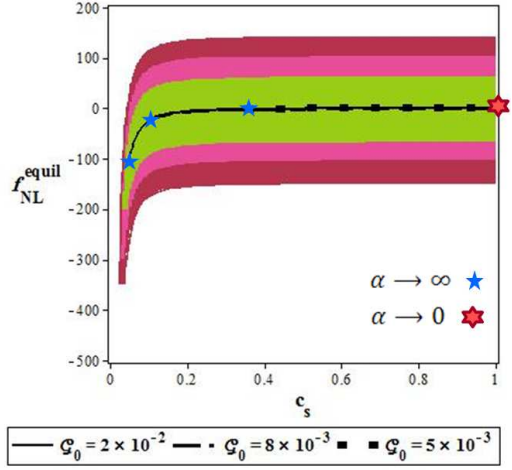}
 \caption{\label{fig2} Equilateral configuration of the
non-Gaussianity versus the sound speed of the perturbations for a
GB-$\alpha$-attractor model in the background of Planck2015 TTT,
EEE, TTE and EET data. When $\alpha\rightarrow 0$ all trajectories
converge to the point corresponding to $f_{NL}\rightarrow0$ and
$c_{s}^{2}\rightarrow 1$. Also, when $\alpha\rightarrow\infty$, the
trajectories tend to the points corresponding to the ones with
$V\sim\frac{\phi^{2}}{2}$ and ${\cal{G}}\sim
{\cal{G}}_{0}\frac{\phi^{2}}{2}$. We have chosen $N = 60$ for the number of e-folds of inflation after the relevant CMB scales cross the horizon.}
\end{figure*}

\section{Reheating}
In this section, we are going to study the reheating era after the
inflation. This process, which during it the universe reheats, helps
us to obtain some additional constraints on the model's parameters.
In this regard, we follow the strategy used in
Refs.~\cite{Dai14,Un15,Co15,Cai15,Ue16} and find some
expressions for $N_{rh}$ and $T_{rh}$. We can write these parameters
in terms of the scalar spectral index $n_{s}$ by which we obtain
constraints on $N_{rh}$ and $T_{rh}$. The number of e-folds
between horizon crossing of the physical scales and the end of
inflation phase is obtained as (see equation \eqref{eq13})
\begin{equation}
\label{eq46} N_{hc}=\ln \left(\frac{a_{f}}{a_{hc}}\right)\simeq
\int_{\phi_{hc}}^{\phi_{f}} \frac{-3H^{2} }{V'+24H^{4}{\cal{G}}'}
d\phi\,,
\end{equation}
with $a_{f}$ as the scale factor at the end of inflation and
$a_{hc}$ as the scale factor at the horizon crossing. Assuming that
during the reheating era, an energy density with the effective
equation of state $\omega_{eff}$ dominates the universe, we can
write $\rho\sim a^{-3(1+\omega_{eff})}$.
We note that with a massive inflaton, $\omega_{eff}$ can be $-1$ (corresponding to the
potential domination) and $+1$ (corresponding to the kinetic
domination). However, at the end of the inflation epoch the value of the effective
equation of state parameter is equal to $-\frac{1}{3}$. On the other hand, at
the beginning of the radiation dominated universe, the value of this
parameter is $\frac{1}{3}$. In this regard, it seems that the
effective equation of state parameter during the reheating epoch is in the
range $-\frac{1}{3}$ to $\frac{1}{3}$. By considering the massive
inflaton, the frequency of the field's oscillations is very larger
than the rate of the expansion at the initial epoch of the
reheating. So, the averaged effective pressure can be neglected and
with a good approximation, the effective equation of state parameter can be
considered to be zero. This means that at the beginning of the reheating
epoch the value of $\omega_{eff}$ is effectively equal to the
equation of state parameter of the matter. Then, by oscillation of the
inflaton field and decaying to other particles, the value of the effective
equation of state parameter increases with time and reaches $\frac{1}{3}$ at
the beginning of the radiation domination era. The authors
of Ref.~\cite{Pod06} have indicated that in the chaotic inflation with a
quartic interaction, the values of $\omega_{eff}$ are in the range
between $0$ and $0.2-0.3$. Regarding that a constant effective
equation of state parameter is allowed by a variety of reheating
scenarios, we consider this parameter as a constant in our
analysis. Now, we obtain the number of e-folds of the reheating era as
\begin{eqnarray}\label{eq47}
N_{rh}=\ln\left(\frac{a_{rh}}{a_{f}}\right)=-\frac{1}{3(1+\omega_{eff})}\ln\left(\frac{\rho_{rh}}{\rho_{f}}\right)\,,
\end{eqnarray}
where the subscript $rh$ demonstrates the parameter in the reheating
epoch. On the other hand, the energy density during inflation is
given by
\begin{eqnarray}\label{eq48}
\rho=\Bigg[1+\frac{\epsilon}{3}-\frac{16\sqrt{2}\kappa^{3}}{9}\sqrt{\epsilon}\,{\cal{G}}'V+\Big(\frac{32}{27}V^{5}
-\frac{64}{9}V^{2}\Big)\kappa^{6}{{\cal{G}}}'^{2}\Bigg]V\,.\nonumber\\
\end{eqnarray}
Now, if we show the value of $k$ at horizon crossing by $k_{hc}$,
then we have
\begin{eqnarray}\label{eq49}
0=\ln\left(\frac{k_{hc}}{a_{hc}H_{hc}}\right)=
\ln\left(\frac{a_{f}}{a_{hc}}\frac{a_{rh}}{a_{f}}\frac{a_{0}}{a_{rh}}\frac{k_{hc}}{a_{0}H_{hc}}\right)\,.
\end{eqnarray}
where $a_{0}$ is the current value of the scale
factor. Equation \eqref{eq49} together with equations
\eqref{eq46} and \eqref{eq47} lead to
\begin{eqnarray}\label{eq50}
N_{hc}+N_{rh}+\ln\left(\frac{a_{0}}{a_{rh}}\right)+\ln\left(\frac{k_{hc}}{a_{0}H_{hc}}\right)=0\,.
\end{eqnarray}
At this point we should find an expression for
$\frac{a_{0}}{a_{rh}}$. The energy density of the reheating epoch
has the following relation with the temperature of this epoch \cite{Co15,Ue16}
\begin{equation}\label{eq51}
\rho_{rh}=\frac{\pi^{2}g_{rh}}{30}T_{rh}^{4}\,,
\end{equation}
where $g_{rh}$ demonstrates the effective number of relativistic
species at the reheating. Also, the conservation of the entropy
gives \cite{Co15,Ue16}
\begin{equation}\label{eq52}
\frac{a_{0}}{a_{rh}}=\left(\frac{43}{11g_{rh}}\right)^{-\frac{1}{3}}\frac{T_{rh}}{T_{0}}\,,
\end{equation}
with $T_{0}$ being the current temperature of the universe. By
using equation \eqref{eq51} and \eqref{eq52} we reach
\begin{eqnarray}\label{eq53}
\frac{a_{0}}{a_{rh}}=\left(\frac{43}{11g_{rh}}\right)^{-\frac{1}{3}}T_{0}^{-1}\left(\frac{\pi^{2}g_{rh}}{30\rho_{rh}}\right)^{-\frac{1}{4}}\,.
\end{eqnarray}
To write $\rho_{rh}$ in terms of $N_{rh}$ from equation
\eqref{eq47}, we should first find $\rho_{f}$. The energy density at
the end of inflation can be obtained by using equation \eqref{eq48}
at the end of inflation (corresponding to $\epsilon=1$). So
\begin{eqnarray}\label{eq54}
\rho_{f}=\Bigg[\frac{4}{3}-\frac{16\sqrt{2}\kappa^{3}}{9}\,{\cal{G}}'_{f}V_{f}+\Big(\frac{32}{27}V_{f}^{5}
-\frac{64}{9}V_{f}^{2}\Big)\kappa^{6}{{\cal{G}}}_{f}'^{2}\Bigg]V_{f}.\nonumber\\
\end{eqnarray}
From equations \eqref{eq47} and \eqref{eq54} we have
\begin{eqnarray}\label{eq55}
\rho_{rh}=\Bigg[\frac{4}{3}-\frac{16\sqrt{2}\kappa^{3}}{9}\,{\cal{G}}'_{f}V_{f}+\Big(\frac{32}{27}V_{f}^{5}
-\frac{64}{9}V_{f}^{2}\Big)\kappa^{6}{{\cal{G}}}_{f}'^{2}\Bigg]V_{f}
\nonumber\\ \times
\exp\Big[-3N_{rh}(1+\omega_{eff})\Big].\hspace{0.7cm}
\end{eqnarray}
By using equation \eqref{eq53} and \eqref{eq55} we get
\begin{eqnarray}\label{eq56}
\ln\left(\frac{a_{0}}{a_{rh}}\right)=-\frac{1}{3}\ln\left(\frac{43}{11g_{rh}}\right)
-\frac{1}{4}\ln\left(\frac{\pi^{2}g_{rh}}{30\rho_{rh}}\right)-\ln T_{0}+\nonumber\\
\frac{1}{4}\ln\left[\bigg(\frac{4}{3}-\frac{16\sqrt{2}\kappa^{3}}{9}\,{\cal{G}}'_{f}V_{f}+\Big(\frac{32}{27}V_{f}^{5}
-\frac{64}{9}V_{f}^{2}\Big)\kappa^{6}{{\cal{G}}}_{f}'^{2}\bigg)V_{f}\right]\nonumber\\-\frac{3}{4}N_{rh}(1+\omega_{eff})\,.\hspace{1cm}
\end{eqnarray}
The value of the Hubble parameter at the horizon crossing ($H_{hc}$)
is obtained from equation \eqref{eq24} as follows
\begin{equation}\label{eq57}
H_{hc}=\sqrt{8{{\cal{A}}_{s}{\cal{W}}_{s}}}\,\pi c_{s}^{\frac{3}{2}}\,.
\end{equation}
Now, using equations \eqref{eq50}, \eqref{eq56} and \eqref{eq57}
gives
\begin{eqnarray}\label{eq58}
N_{rh}=\frac{4}{1-3\omega_{eff}}\Bigg[-N_{hc}-\ln\Big(\frac{k_{hc}}{a_{0}T_{0}}\Big)-\frac{1}{4}\ln\Big(\frac{40}{\pi^{2}g_{rh}}\Big)\nonumber\\
-\frac{1}{3}\ln\Big(\frac{11g_{rh}}{43}\Big)+\frac{1}{2}\ln\Big(8\pi^{2}{\cal{A}}_{s}{\cal{W}}_{s} c_{s}^{3}\Big)
-\frac{1}{4}\ln\bigg(\Big[\frac{4}{3}\nonumber\\-\frac{16\sqrt{2}\kappa^{3}}{9}\,{\cal{G}}'_{f}V_{f}+\Big(\frac{32}{27}V_{f}^{5}
-\frac{64}{9}V_{f}^{2}\Big)\kappa^{6}{{\cal{G}}}_{f}'^{2}\Big]V_{f}\bigg)\Bigg].\nonumber\\
\end{eqnarray}
Also, from equations \eqref{eq47}, \eqref{eq52} and \eqref{eq54} we
have
\begin{eqnarray}\label{eq59}
T_{rh}=\bigg(\frac{30}{\pi^{2}g_{rh}}\bigg)^{\frac{1}{4}}\hspace{5.5cm}\nonumber\\
\times
\Bigg[\frac{4}{3}-\frac{16\sqrt{2}\kappa^{3}}{9}\,{\cal{G}}'_{f}V_{f}+\Big(\frac{32}{27}V_{f}^{5}
-\frac{64}{9}V_{f}^{2}\Big)\kappa^{6}{{\cal{G}}}_{f}'^{2}\Bigg]^{\frac{1}{4}}
V_{f}^{\frac{1}{4}} \nonumber\\ \times
\exp\Big[-\frac{3}{4}N_{rh}(1+\omega_{eff})\Big]\,.\hspace{1cm}
\end{eqnarray}
By substituting equations \eqref{eq1} and \eqref{eq3} in equations
\eqref{eq58} and \eqref{eq59}, we obtain the e-folds number and
temperature during reheating in terms of the model's parameters,
especially $\phi_{hc}$. On the other hand, $\phi_{hc}$ is related to
the scalar spectral index, as is seen from equation \eqref{eq33} at
the horizon crossing. So, it is possible to study the number of
e-folds and temperature at the reheating versus the scalar spectral
index. In our numerical analysis, we have adopted some sample values of $\omega_{eff}$ which are in
the range between $-\frac{1}{3}$ and $1$. We note that the values $-\frac{1}{3}$
and $1$ are the very conservative allowed values. The results are shown in figures 3 and 4. Note that in plotting the
figures and finding the constraints, we have used the value of the
scalar spectral index as $n_{s}=0.9652 \pm 0.0047$ from Planck2015
TT,TE,EE+lowP. As the figures show, for each class of the model's
parameters values, there is a maximum value for the number of
e-folds which is consistent with observational data. In table II we
have summarized the results for some values of the model's
parameters. Also, the figures show that there is a point that all
curves in the plots originate from it. This point is corresponding
to $N_{rh}=0$ as instantaneous reheating. Note that, for GB coupling
constant of the order of $10^{-2}$ or more and $\alpha$ of the order
of 10, the instantaneous reheating is disfavored by Planck2015
observational data. We have found that, for all values of the
parameter $\alpha$, as the GB coupling constant ${\cal{G}}_{0}$
becomes smaller, the larger values of the e-folds number are more
viable. Similarly, we can study the temperature in reheating phase
numerically. For each class of the model's parameter values there
is a minimum temperature which is consistent with observational data
(see table III). It is turned out that, for a given value of
$\alpha$, as ${\cal{G}}_{0}$ becomes smaller, a wider range of the
reheating temperature is viable.

It should be noticed that, for
$\omega_{eff}=\frac{1}{3}$ we can't use equation \eqref{eq58}.
Actually even we repeat the analysis from equation \eqref{eq47} with
$\omega_{eff}=\frac{1}{3}$, we can't find an explicit expression for
number of e-folds and temperature. However, it is possible to almost
predict the value of the scalar spectral index. The curve for
$\omega_{eff}=\frac{1}{3}$ in the plots can be a vertical line that
crosses the instantaneous reheating point (see~\cite{Un15,Co15}).

\begin{table*}
\caption{\label{tab:2} The ranges of the number of e-folds parameter
at reheating that are consistent with observational data.}
\begin{tabular}{cccccccc}
\\ \hline \hline&$\alpha=0.1\,,\,{\cal{G}}_{0}=8\times 10^{-3}$&&$\alpha=0.1\,,\,{\cal{G}}_{0}=2\times 10^{-2}$&&
$\alpha=10\,,\,{\cal{G}}_{0}=8\times 10^{-3}$&&$\alpha=10\,,\,{\cal{G}}_{0}=2\times 10^{-2}$\\ \hline\\
Planck\,,\,$\omega=-\frac{1}{3}$&  $N_{rh}\leq15$ &&$N_{rh}\leq6$&&
$N_{rh}\leq13$ &&--------\\\\Planck\,,\,$\omega=0$& $N_{rh}\leq24$
&&$N_{rh}\leq10$&& $N_{rh}\leq17$
&&--------\\\\Planck\,,\,$\omega=\frac{1}{6}$& $N_{rh}\leq78$
&&$N_{rh}\leq36$&& $N_{rh}\leq41$
&&--------\\\\Planck\,,\,$\omega=1$& $N_{rh}\leq50$
&&$N_{rh}\leq80$&& $N_{rh}\leq68$
&&$N_{rh}\geq 4$\\\\
\hline
\end{tabular}
\end{table*}

\begin{figure*}
\flushleft\leftskip0em{
\includegraphics[width=.38\textwidth,origin=c,angle=0]{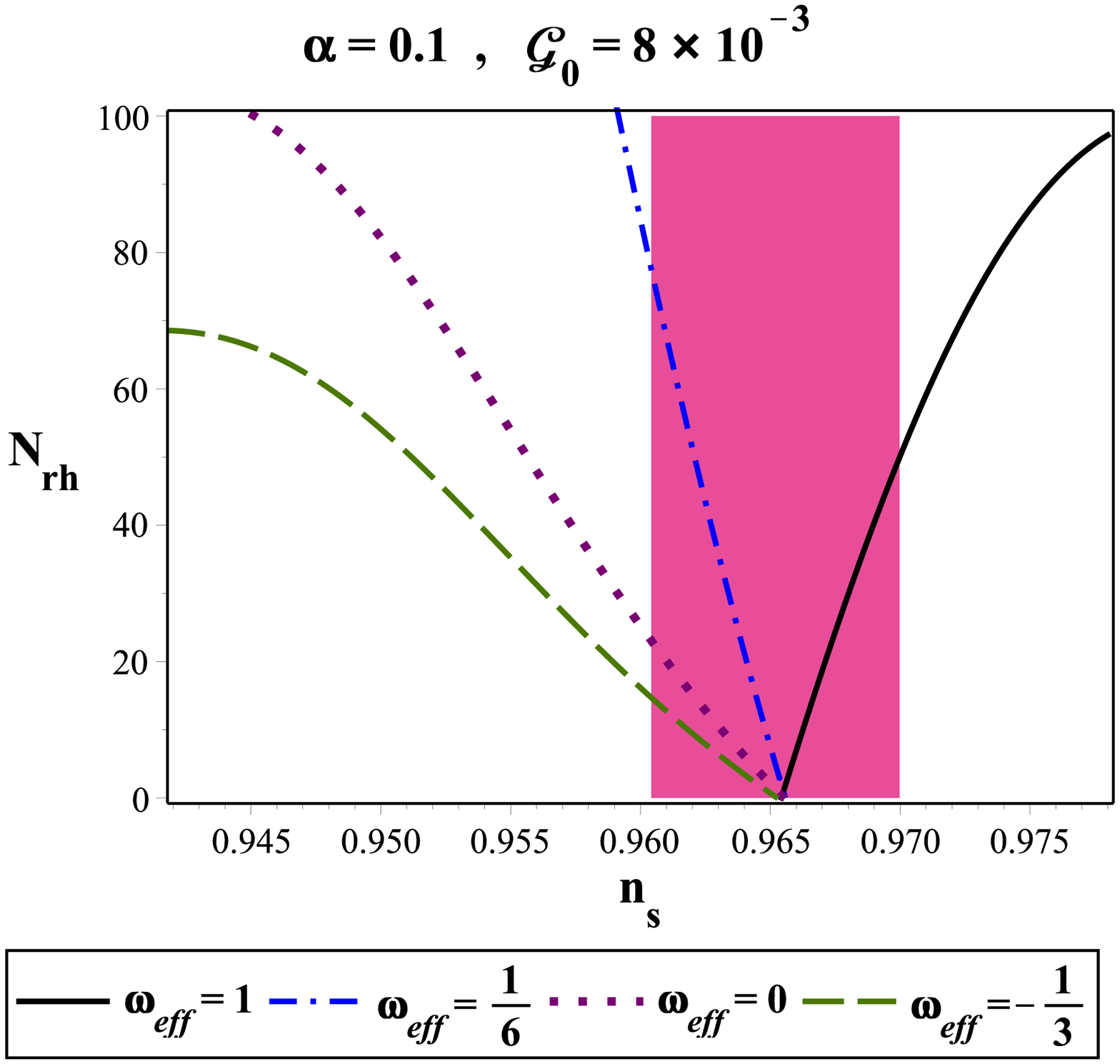}
\hspace{3cm}
\includegraphics[width=.38\textwidth,origin=c,angle=0]{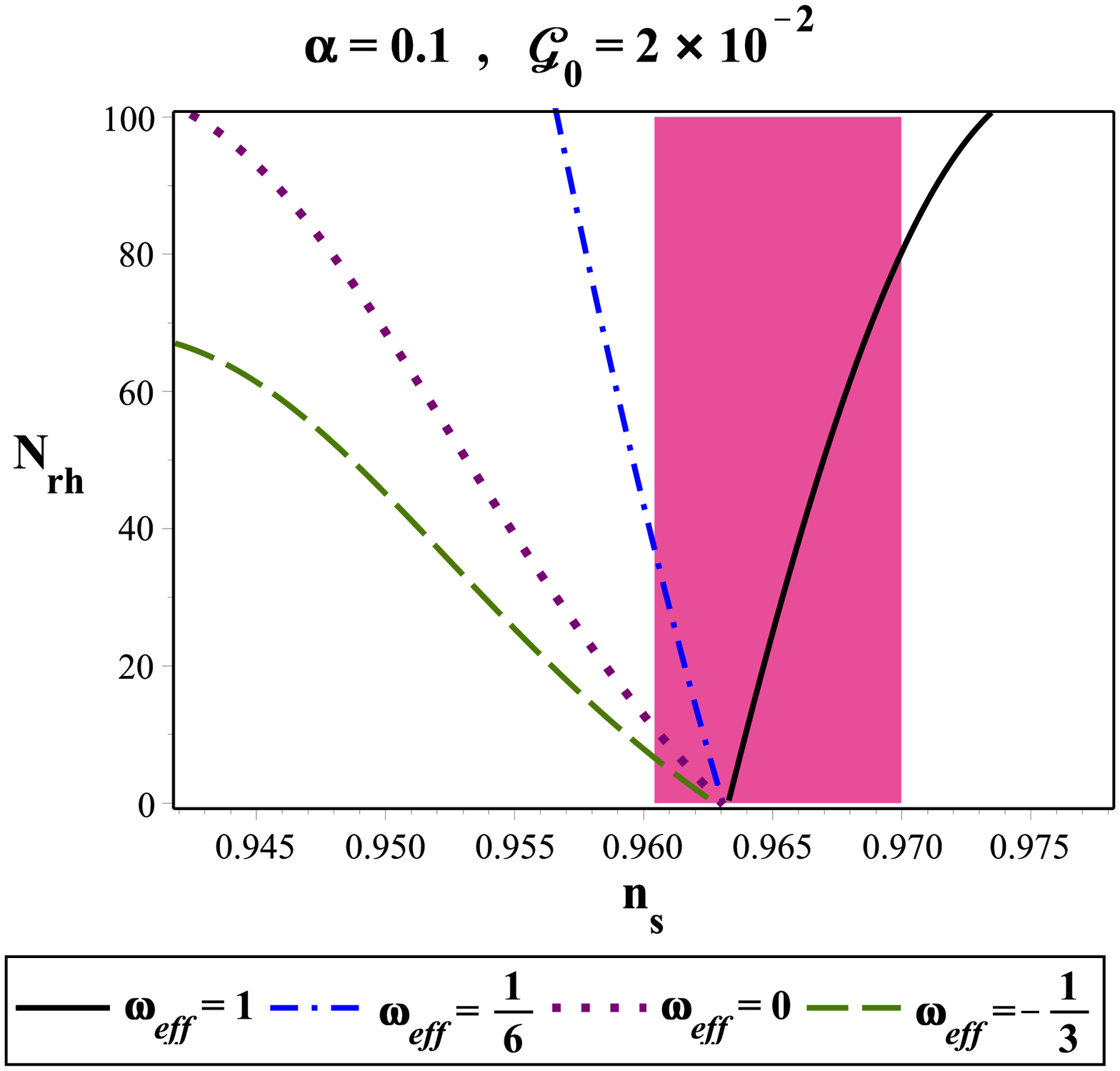}\\
\vspace{1cm}
\includegraphics[width=.38\textwidth,origin=c,angle=0]{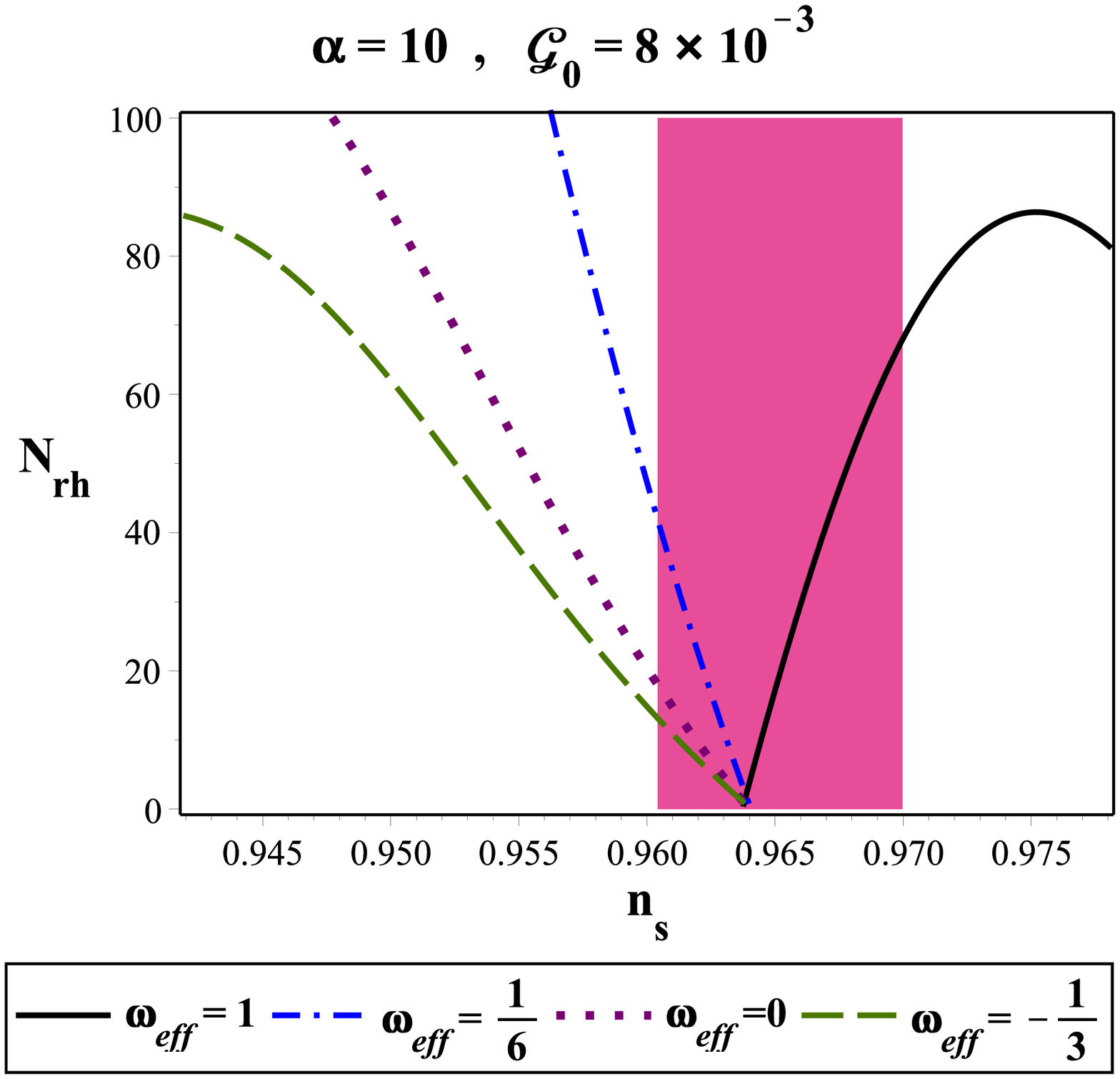}
\hspace{3cm}
\includegraphics[width=.38\textwidth,origin=c,angle=0]{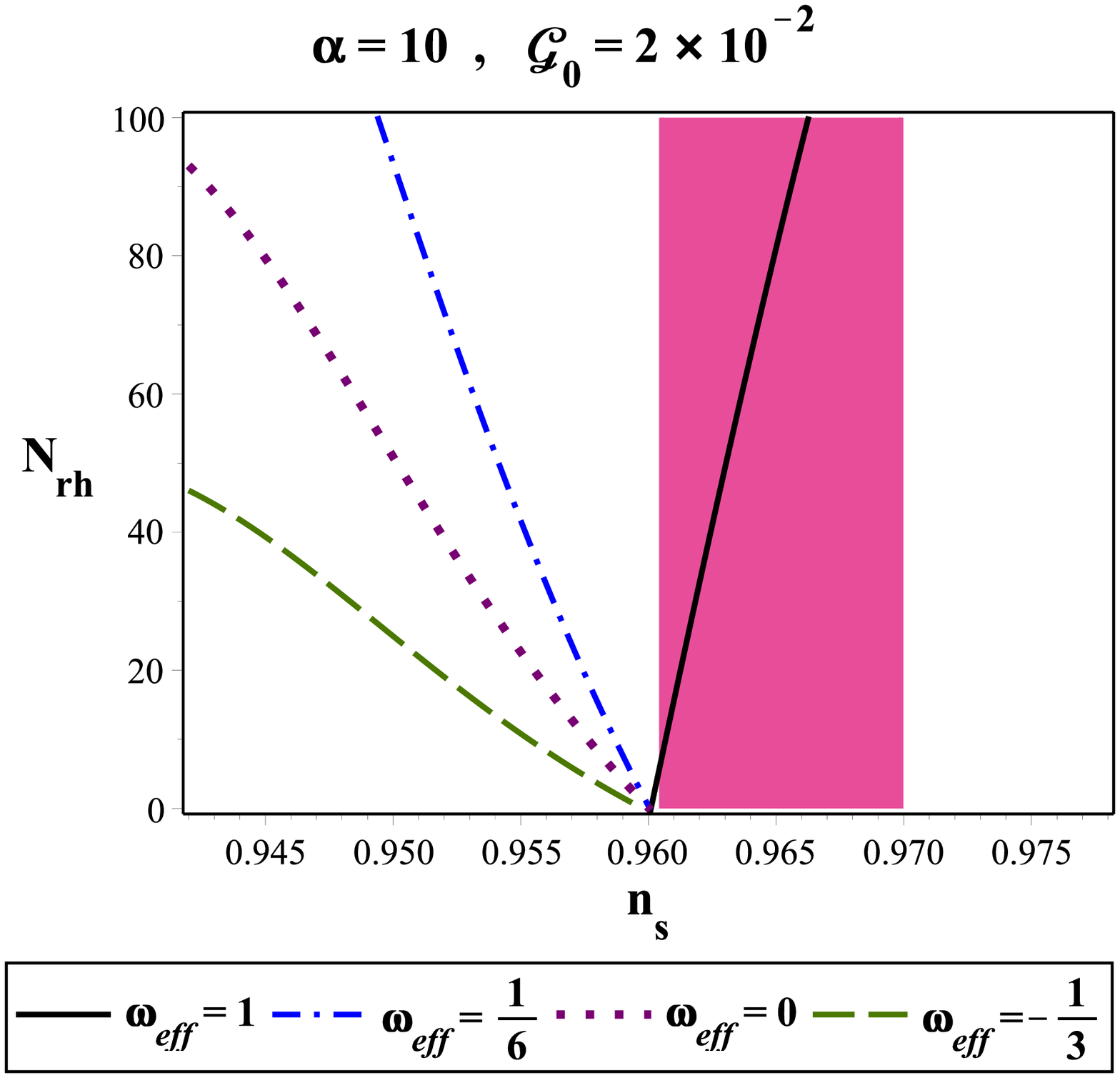}
} \caption{\label{fig3} Number of e-folds in reheating epoch versus
the scalar spectral index for an E-model GB-$\alpha$-attractor
inflation, with $n=1$. The magenta region shows the values of the
scalar spectral index released by Planck2015 data.}
\end{figure*}

\begin{table*}
\caption{\label{tab:3} The ranges of the temperature at reheating
which are consistent with observational data.}
\begin{tabular}{cccccccc}
\\ \hline \hline&$\alpha=0.1\,,\,{\cal{G}}_{0}=8\times 10^{-3}$&&$\alpha=0.1\,,\,{\cal{G}}_{0}=2\times 10^{-2}$&&
$\alpha=10\,,\,{\cal{G}}_{0}=8\times 10^{-3}$&&$\alpha=10\,,\,{\cal{G}}_{0}=2\times 10^{-2}$\\ \hline\\
Planck\,,\,$\omega=-\frac{1}{3}$&
$\log_{10}\left(\frac{T_{rh}}{GeV}\right)\geq13.3$
&&$\log_{10}\left(\frac{T_{rh}}{GeV}\right)\geq13.65$&&
$\log_{10}\left(\frac{T_{rh}}{GeV}\right)\geq12.71$
&&--------\\\\Planck\,,\,$\omega=0$&
$\log_{10}\left(\frac{T_{rh}}{GeV}\right)\geq10.67$
&&$\log_{10}\left(\frac{T_{rh}}{GeV}\right)\geq13.29$&&
$\log_{10}\left(\frac{T_{rh}}{GeV}\right)\geq11.38$
&&--------\\\\Planck\,,\,$\omega=\frac{1}{6}$&
$\log_{10}\left(\frac{T_{rh}}{GeV}\right)\geq-0.16$
&&$\log_{10}\left(\frac{T_{rh}}{GeV}\right)\geq5.71$&&
$\log_{10}\left(\frac{T_{rh}}{GeV}\right)\geq5.43$
&&--------\\\\Planck\,,\,$\omega=1$&
$\log_{10}\left(\frac{T_{rh}}{GeV}\right)\geq4.39$
&&$\log_{10}\left(\frac{T_{rh}}{GeV}\right)\geq1.18$&&
$\log_{10}\left(\frac{T_{rh}}{GeV}\right)\geq2.95$
&&$\log_{10}\left(\frac{T_{rh}}{GeV}\right)\leq 13.94$\\\\
\hline
\end{tabular}
\end{table*}

\begin{figure*}
\flushleft\leftskip0em{
\includegraphics[width=.38\textwidth,origin=c,angle=0]{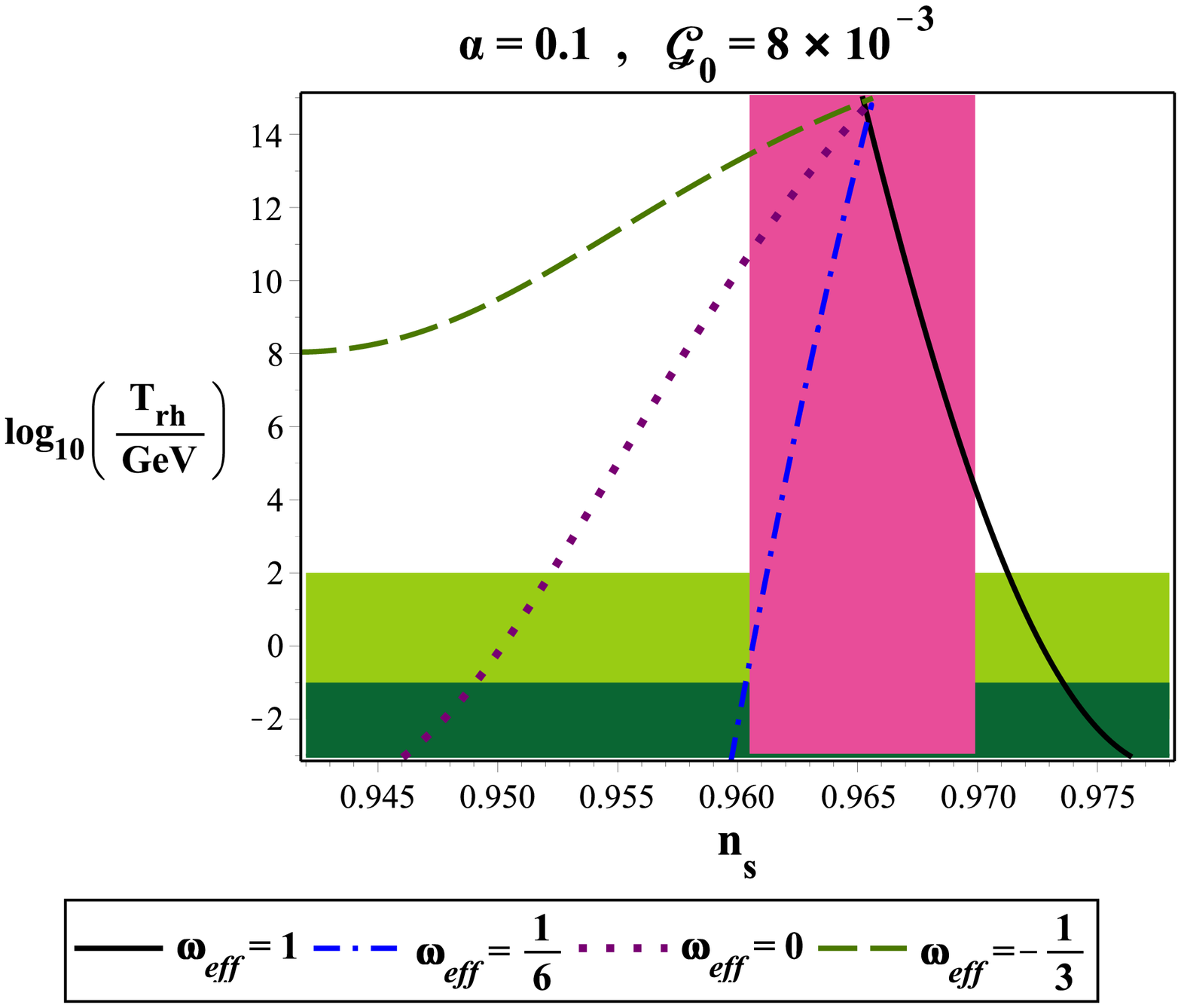}
\hspace{3cm}
\includegraphics[width=.38\textwidth,origin=c,angle=0]{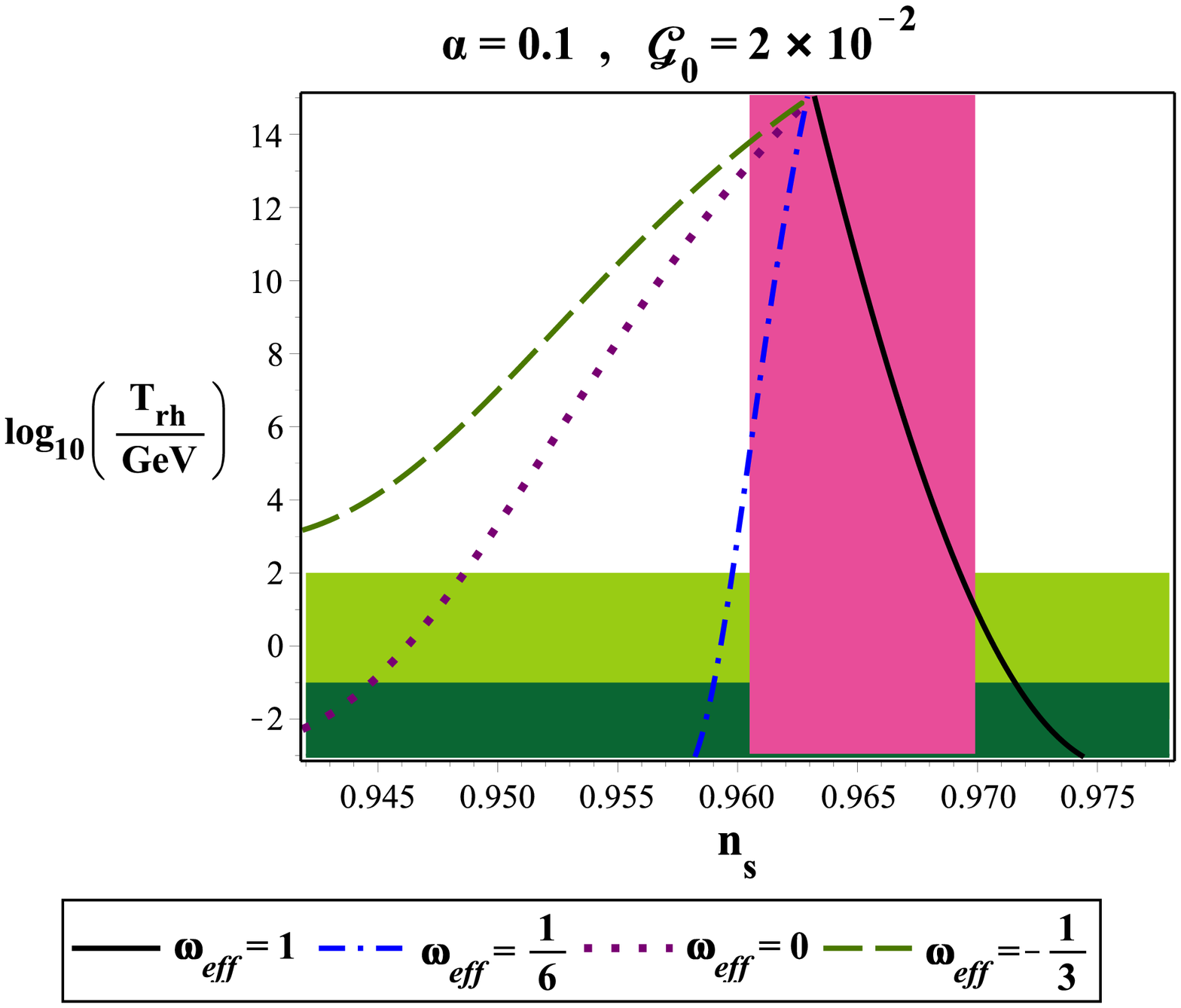}\\
\vspace{1cm}
\includegraphics[width=.38\textwidth,origin=c,angle=0]{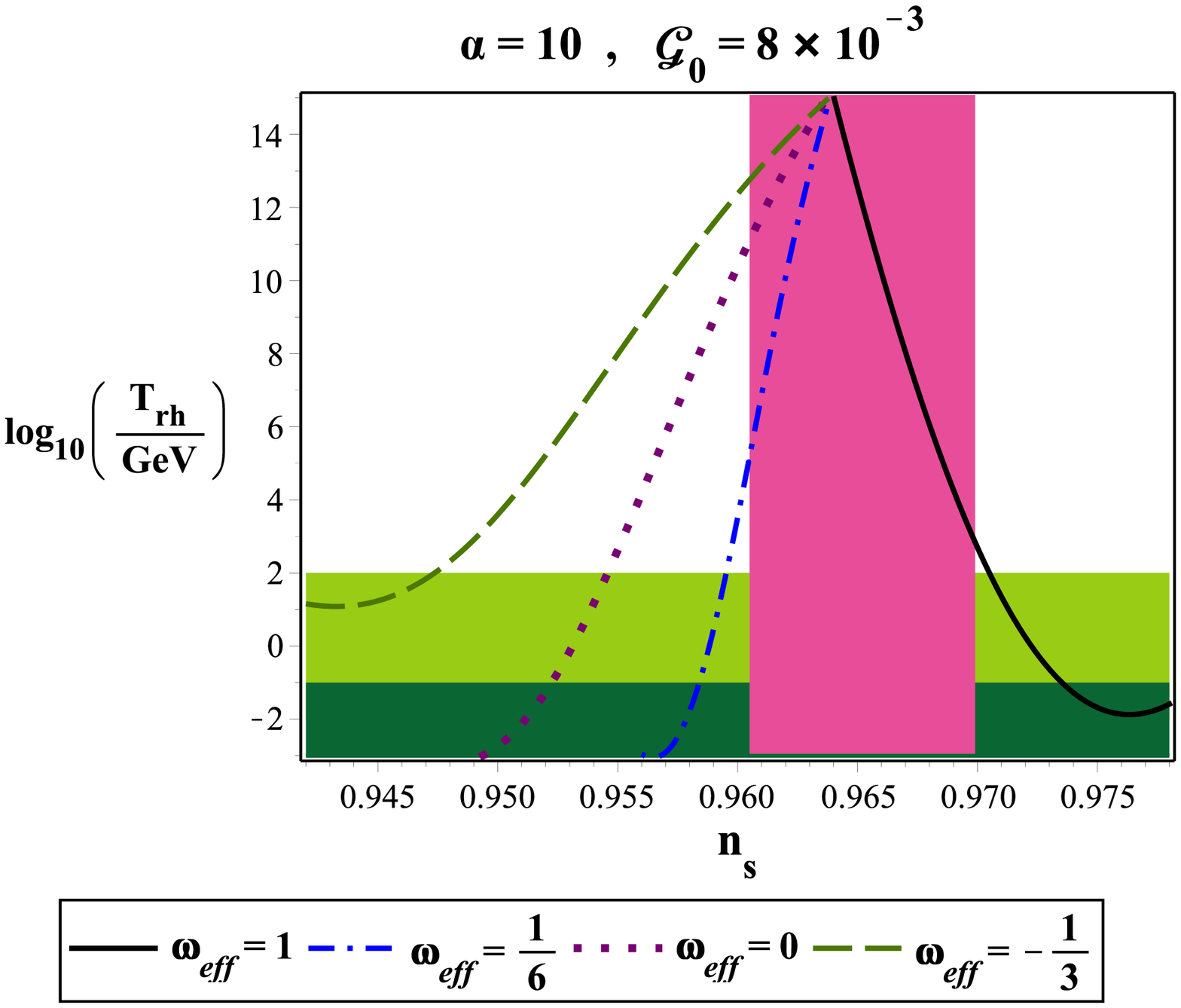}
\hspace{3cm}
\includegraphics[width=.38\textwidth,origin=c,angle=0]{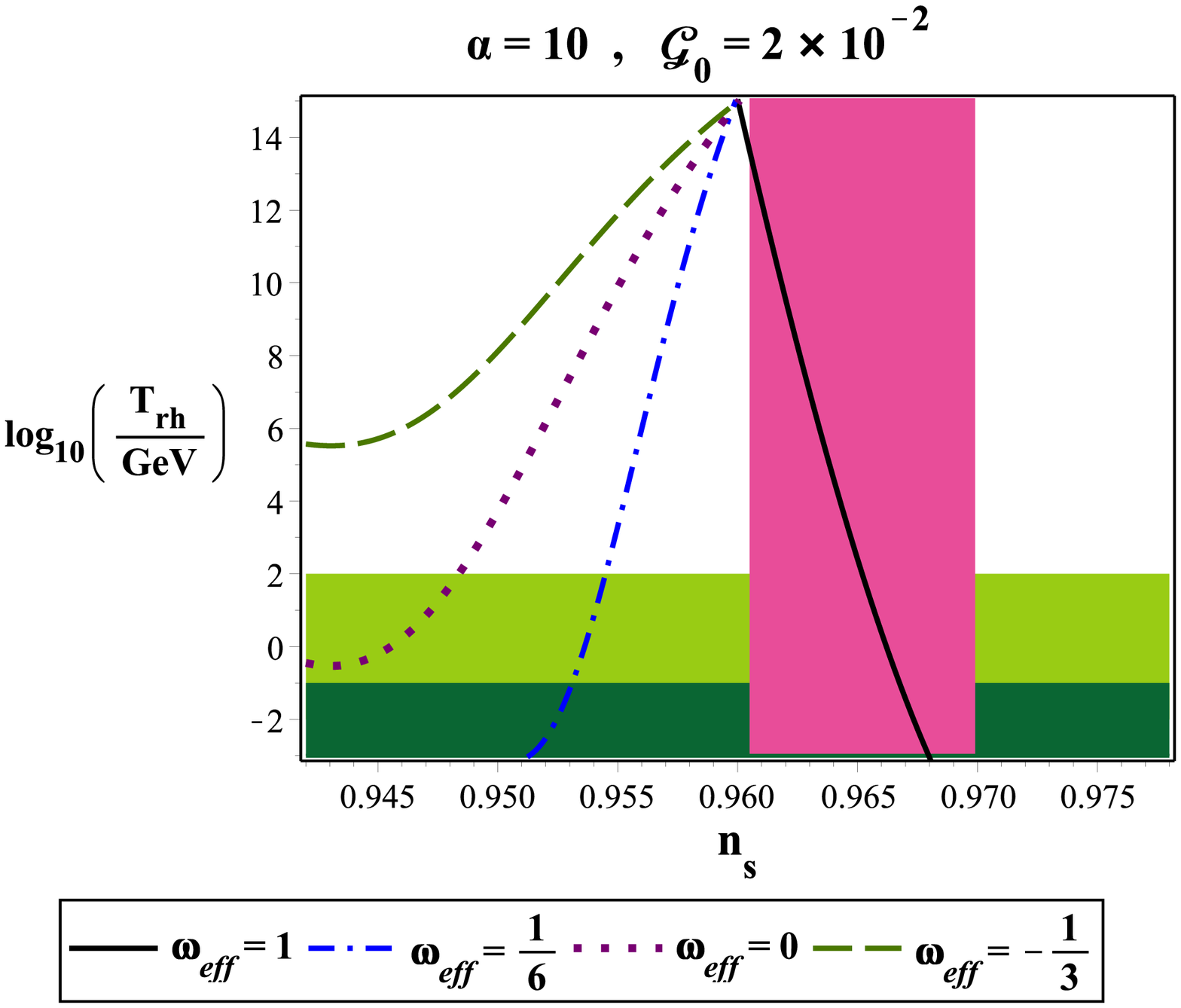}
} \caption{\label{fig4} Temperature in reheating epoch versus the
scalar spectral index for an E-model GB-$\alpha$-attractor
inflation with $n=1$. The light green region shows the temperatures
below the electroweak scale, $T<100GeV$ and the dark green region
demonstrates the temperatures below the big bang nucleosynthesis
scale, $T<10MeV$.}
\end{figure*}

\section{Summary}

In this paper, we have considered a non-minimal Gauss-Bonnet
inflation in the spirit of attractor scenario. Motivated by the
$\alpha$-attractor models, we have adopted the potential and
Gauss-Bonnet coupling function to be the same as the E-model
potential. In this regard, in the small $\alpha$ limit there is an
attractor point and in the large $\alpha$ limit we obtain
$V\sim\phi^{2n}$ and ${\cal{G}}\sim\phi^{2n}$. We have analyzed the
perturbations in this setup and obtained the important perturbation
parameters, such as the scalar spectral index and tensor-to-scalar
ratio. To test the observational viability of this
GB-$\alpha$-attractor model, we have studied $r$ and $n_{s}$
numerically and compared the results with Planck2015 data for $n=1$.
As expected, in small $\alpha$ limit the tensor-to-scalar ratio for
all values of ${\cal{G}}_{0}$ tends to zero and the scalar spectral
index tends to 0.967. In the large $\alpha$ limit the model is
corresponding to the model with $V\sim\phi^{2}$ and
${\cal{G}}\sim\phi^{2}$. Our study shows that the
GB-$\alpha$-attractor model in some domains of the model's
parameter space is consistent with observational data. The
non-Gaussian feature of the primordial perturbations also, have been
studied in this setup and the amplitude of the non-Gaussianity has
been obtained. We have obtained $f_{NL}$ in the equilateral
configuration of the non-Gaussianity characterized by
$k_{1}=k_{2}=k_{3}$. By studying the behavior of $f_{NL}^{equil}$
versus the sound speed, we have found that in small limit of
$\alpha$ the amplitude of non-Gaussianity tends to zero
corresponding to $c_{s}\rightarrow 1$. However, in the large limit of
$\alpha$, depending on the GB coupling constant ${\cal{G}}_{0}$, the
sound speed can be small and $f_{NL}^{equil}$ can be large. So, in
this GB-$\alpha$-attractor model there is a set of model parameters
which makes $f_{NL}^{equil}$ large enough to be detected in future
surveys. The reheating era after the inflation epoch has been
explored in this GB-$\alpha$-attractor model to obtain some more
constraints on the model's parameters. By assuming the domination of
an energy density with an effective equation of state and
considering its relation with the temperature, we have obtained the
number of e-folds and temperature of the reheating in terms of the
model's parameters. After that, regarding to the relation of the
scalar field at horizon crossing with the scalar spectral index, we
have performed the numerical analysis on the behavior of $N_{rh}$
and $T_{rh}$ versus $n_{s}$. The results have been compared to
Planck2015 constraints on the scalar spectral index. In this regard,
we have obtained some constraints on the e-folds number and
temperature of the reheating which are corresponding to the
observationally viable values of the scalar spectral index. As an
important result, we found that, for all values of the parameter
$\alpha$, as the GB coupling constant reduces, the larger values of
the e-folds number are more viable observationally.\\

{\bf Acknowledgement}\\
We are grateful to the referee for very insightful comments that improved the work considerably.\\

\end{document}